 \documentclass[final,5p,times,singlecolumn]{elsarticle}

\usepackage{graphicx}
 
 \usepackage{epsfig}
\usepackage{amssymb}
  \usepackage{amsthm}

\begin{document}

\begin{frontmatter}



\title{Cosmic ray modulation studies with Lead-Free Gulmarg Neutron Monitor}

 
\author []{M. A. Darzi}
\ead{darzi.mushtaq.ahmad@gmail.com}
\author[]{P. M. Ishtiaq}
\author[]{T. A. Mir}
\author[]{S. Mufti}
\author[]{G. N. Shah$^{*}$}
\address{Nuclear Research Laboratory, Astrophysical Sciences Division, Bhabha Atomic Research Centre\\ Srinagar - 190006, Kashmir, India} 

\address{$^{*}$52-Hamzah Colony Bemina, Srinagar - 190012, Kashmir, India} 


\begin{abstract}
A lead-free neutron monitor operating at High Altitude Research Laboratory (HARL), Gulmarg optimized for detecting 2.45 MeV neutron bursts produced during the atmospheric lightning discharges is also concurrently used for studying background neutron component present in the atmosphere. These background neutrons are produced due to the interaction of primary cosmic rays with the atmospheric constituents. In order to study and extract the information about the yield of the neutron production during  transient atmospheric lightning discharges, the system is continuously operated to monitor and record the  cosmic ray produced background secondary neutrons in the atmosphere. The data analysis of the background neutrons recorded by Lead-Free Gulmarg Neutron Monitor (LFGNM) has convincingly established that the modulation effects due to solar activity phenomena compare very well with those monitored by the worldwide IGY or NM64 type neutron monitors which have optimum energy response relatively towards the higher 
energy regime of the cosmic rays. The data has revealed various types of modulation phenomena like diurnal variation, Forbush decrease etc during its entire operational period. However, a new kind of a periodic/seasonal variation pattern is also revealed in the data from September 2007 to September 2012, which is seen to be significantly consistent with the data recorded by Emilio Segre observatory, Israel (ESOI) Neutron Monitor. Interestingly, both these neutron monitors have comparable latitude and altitude. However, the same type of consistency  is not observed in the data recorded by the other conventional neutron monitors operating across the globe.

\end{abstract}


\end{frontmatter}


\section{Introduction}

\label{1}
A host of secondaries are produced on interaction of the primary cosmic rays with the earth's atmosphere of which neutrons form a dominant component  \cite{Hess1961,Phillips2001}. Ground based monitors are widely used for observation of these  secondary nucleonic components present in the earth's atmosphere. These observations serve to indirectly study the different characteristics of the causative phenomena which are responsible for the modulation of the cosmic rays. The conventional IGY \cite{Simpson1953, Simpson1957} and NM64 \cite{Carmichaell968} neutron monitors have been extensively utilized for studying the changes in the cosmic ray secondary neutron component and thereby, have contributed in establishing the existence of both short-term variation (Forbush Decreases (FD) and Solar Energetic Particle (SEP) events) and long-term variations (11-year and 22-year variation) in the intensity of the galactic cosmic rays incident on the earth. These ground based neutron monitors are distributed world over to 
cover a wide range of geomagnetic cut-off rigidities. The count rate of these monitors has a high contribution from the secondary high-energy neutrons and to some extent by protons that interact with the constituent lead producing spallation and thereby enhancing the count rate in these monitors \cite{Hughes1966}. However, Shah et al.\cite{Shah2005,Shah2007} has exploited the features of a neutron monitor with out lead -LFGNM- to study modulation effects. LFGNM is the modified configuration of the conventional IGY neutron monitor earlier operating at the same location. Due to the complete removal of the constituent lead and other major changes in the conventional IGY neutron monitor design particularly the reduction of thickness of upper paraffin reflector from 28 cm to 8 cm only, LFGNM is optimized primarily for detecting 2.45 MeV neutrons produced during the natural lightning discharges in the atmosphere \cite{Shah1985}. Consquently, the threshold of the energy response in LFGNM has been reduced relatively 
towards the lower energy regime of the secondary neutrons \cite{Shah2010,Mufti2011}. To study and extract the information about the yield of neutron production during transient lightning events from the sea of background neutrons present in the atmosphere, LFGNM is operated continuously to record this neutron background. Therefore, LFGNM serves the dual purpose of monitoring neutrons produced during  transient lightning discharges as well as the background cosmic ray secondary neutrons always present in the atmosphere. The background cosmic ray neutron data obtained has been analyzed to see the response of LFGNM to changes in the cosmic ray component vis-a-vis conventional neutron monitors. We present here the results showing both short-term and long-term variation in the background cosmic ray neutron data recorded by LFGNM. We observe that the response of LFGNM to Forbush Decrease events due to transient solar activity is in conformity with the response obtained with conventional neutron monitors. However, 
a 
new type of periodic/seasonal variation has been recorded by LFGNM during the period from September 2007 to September 2012. While exploring the existence of such a kind of variation in the data recorded by the other monitors during the same period, the authors after analysis and comparison of the data obtained from these neutron monitors have found that a similar kind of variation is exhibited only by ESOI neutron monitor which is a neutron monitor having 6NM64 configuration containing constituent lead \cite{Dorman1999}. The lead is chosen for producing evaporation neutrons due to its interaction with the incident energetic nucleons to enhance the detection probability and thereby improving the data statistics in these kind of neutron monitors. In addition, lead has relatively low absorption cross-section for thermal neutrons. Significantly, it is seen that the threshold energy response of an NM64 neutron monitor configuration is consistent with the threshold energy response of LFGNM. This may be  due to the 
fact that the mass thickness of upper moderator of LFGNM is almost same as the mass thickness of the upper reflector of ESOI neutron monitor  and the relatively smaller thickness (8 cm in case of LFGNM and 7.5 cm in case of ESOI neutron monitor) of moderator/reflector renders both these neutron monitors more susceptible to the variations of the environmentally produced neutrons as compared to the conventional IGY neutron monitor \cite{Stoker2000}. The threshold energy response of  IGY neutron monitor to the incident secondary cosmic ray neutrons is $\geqq$ 50 MeV and for  protons it is $\geqq$ 180 MeV whereas in the case of LFGNM and ESOI neutron monitor the threshold energy response starts from a 0.01 eV energy  for incident neutrons \cite{Hess1961,Hughes1966,Ballerini1969}. Therefore, the observational features of LFGNM particularly the kind of a response to the incident secondary cosmic ray neutrons makes it a very good complementary tool for studying variations in the cosmic ray intensity relatively 
towards the lower energy regime.\
\section{Data analysis and Results}
\label{2}
  LFGNM records intensity of the neutrons present in the atmosphere. However, the meteorological parameters like the temperature, the wind and the atmospheric pressure of the neutron monitor station affects its observed count rate. Therefore for extracting any meaningful information from observed LFGNM data, it is corrected for the continuous changes occurring in these atmospheric parameters. The temperature effects are generally observed to be small and negligible for the nucleonic  component \cite{Dorman1974,Iucci1999,Duldig2000}. However, the temperature sensitivity of the instruments employed in the neutron monitors is a significant issue, and has been subject of discussion in many scholarly works \cite{Evenson2005,Kruger2008}. The temperature sensitivity of a neutron monitor is given in terms of its  combined temperature coefficient of its constituent components and for NM64 neutron monitor configuration, by simulation, it has been found to be 0.018 $\pm$ 0.006\%  per $^\circ$C \cite{Kruger2008}. 
Assuming and 
applying the same temperature coefficient figure of 0.018 $\pm$ 0.006 \% per $^\circ$C to LFGNM data, it is found that over a temperature variation of -12 $^\circ$C to +24  $^\circ$C (an absolute temperature variation of 36 $^\circ$C observed) during the period from core of winter to the peak of summer respectively at HARL, Gulmarg, the percentage variation in the counts has been found to be  \textless 1\%. Hence the count rate of the LFGNM is not corrected for atmospheric temperature variations. Similarly, the wind correction for LFGNM data is safely discarded \cite{Lockwood1957} because Gulmarg is not prone to high speed winds and the average wind speed recorded at Gulmarg has not been more than 1 $ms^{-1}$. However, the count rate of the neutron monitors is drastically affected by the changes in the atmospheric pressure and is always anti-correlating with the station atmospheric pressure. Therefore the pressure induced effects of LFGNM data need to be evaluated. We have adopted the method of successive 
differences \cite{Lapointe1962,Batchlet1965,Griffiths1966} for computing the pressure coefficient of LFGNM. The pressure corrected neutron count of LFGNM is given by the following expression
 
                   \begin{equation}       
                          N = N_{0} e^{-\beta(P_{obs}-P_{avg})} 
                          \end{equation}  
                          \begin{figure}
 \hspace{0pt}
\centerline{\epsfig{file=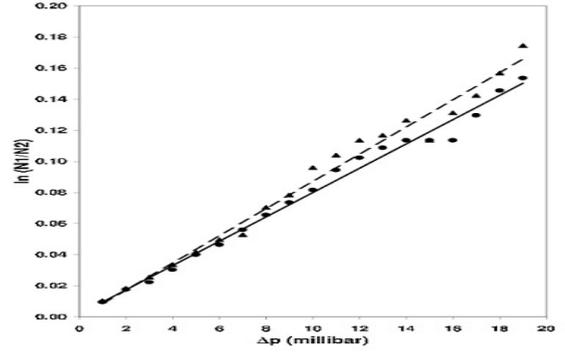,width=9cm, height=5.75cm}}
\caption{Regression lines of ln(N$_{1}$/N$_{2}$) on $\triangle$p i.e., the change in atmospheric pressure, fitted to increasing pressure (triangles) and decreasing pressure (dots) data subsets of the major pressure troughs. N$_{1}$ and N$_{2}$ represent neutron monitor counting rates corresponding to pressure P$_{1}$ and P$_{2}$ such that $\triangle$p = P$_{1}$-P$_{2}$. \label{Figure 1.}} 
\end{figure}
 where P$_{obs}$ is the observed pressure and P$_{avg}$ is the average pressure at the LFGNM site. The quantity $\beta$ is called the barometric coefficient and its value reflects the change in count rate with respect to the changes in atmospheric pressure of the station. Two types of data sets have been employed for computing the pressure coefficient $\beta$ for LFGNM. The first data set comprises of fourteen major pressure troughs registered at the station. Each of these troughs persisted for several days and the pressure recorded in these troughs varied from 725 to 744 millibars, therefore providing a pressure range of 19 millibars. Each pressure trough provides us with two data subsets: one corresponding to the decreasing values of pressure within the trough, and the other corresponding to the increasing values of the pressure \cite{Shah1985,Shah2010}. We present in figure 1 the regression lines fitted to the decreasing (dots) and increasing (triangles) pressure data subsets shown by the full and  broken 
lines respectively. The slope of the regression lines yield an average  $\beta$  value  of -0.845$\pm$ 0.020 \% per millibar with the correlation coefficient as high as -0.98.

\begin{figure} 
\hspace{-20pt}
\centerline{\epsfig{file=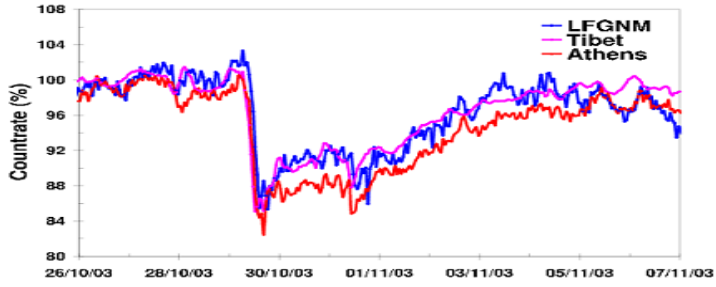,width=5.5cm, height=11.0cm, angle=270}}
\vspace{-20pt}
\caption{Comparison of the Forbush Decrease event of October-November 2003 recorded by LFGNM with Athens and Tibet neutron monitors.\label{Figure 2.}}
\end{figure}
\begin{figure}
\vspace{-20pt}
\hspace{-20pt}
\centerline{\epsfig{file=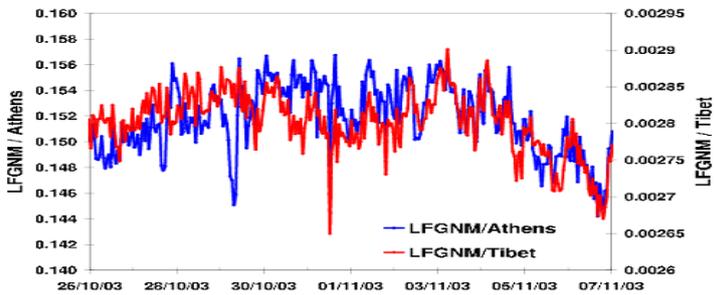,width=5.5cm, height=11cm, angle=270}}
\vspace{-20pt}
\caption{Ratio of LFGNM/Athens and LFGNM/Tibet.\label{Figure 3.}}
\end{figure}
\begin{figure}
\hspace{-20pt}
\centerline{\epsfig{file=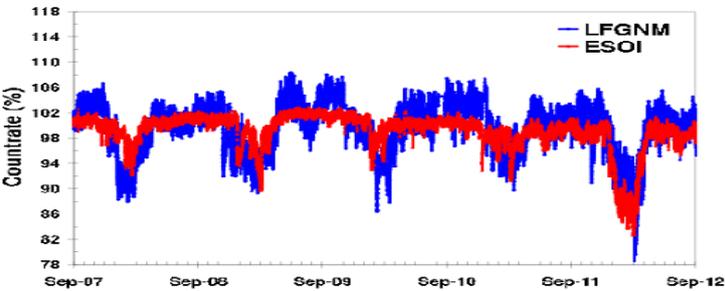,width=5.5cm, height=11cm, angle=270}}
\vspace{-20pt}
\caption{Hourly data profile of LFGNM and  ESOI neutron monitor from September 2007 to September 2012.\label{Figure 4.}}
\end{figure}
\begin{figure}
\hspace{-20pt}
\centerline{\epsfig{file=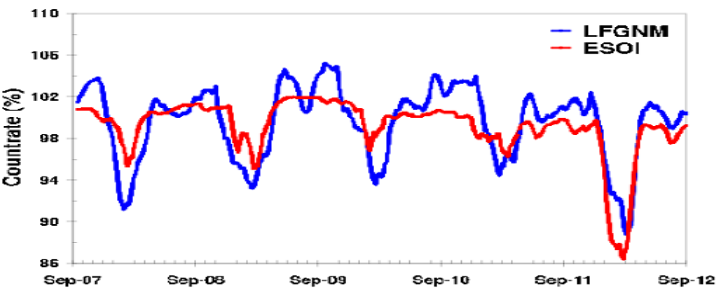,width=5.5cm, height=11cm, angle=270}}
\vspace{-20pt}
 \caption{27-day moving average data profile of LFGNM and  ESOI neutron monitor from September 2007 to September 2012.\label{Figure 5.}}
\end{figure}
The pressure corrected data of LFGNM during the period of the Forbush Decrease event of $29^{th}$, October 2003 has been analyzed to look for the response of the monitor to this transient event. The $29^{th}$, October 2003 Forbush Decrease event has onset time of 09:51 UT and is attributed to X17.2/4B class solar flare occurring on 28th October 2003 \cite{Gopal2005}. The percentage count rate of LFGNM  for the period of October 26th to November 5, 2003 is presented along with that of Athens and Tibet neutron monitors in figure 2. A decrease of 16\% followed by a characteristic  slow recovery is observed in the response of LFGNM, which are typical of a Forbush Decrease induced modulation effect. Similarly Athens and Tibet neutron monitors have recorded comparable decreases for this  event. The 100\% count rate has been taken as an average of 72 hours of counts before the  occurrence of the onset time. A similarity is observed both in the profile of the variation and the percentage amplitude 
decrease recorded by these monitors, indicating that a lead-free neutron monitor is as effective in studying the short-term transient modulation effects though having optimum response relatively towards the lower energy regime of the cosmic rays. Furthermore to quantitatively depict this transient variation, the ratio LFGNM/Athens and LFGNM/Tibet are plotted in figure 3.  As is clear from the figure the respective ratios are seen to be consistent during transient as well as quite sun conditions of this event period.  
\newline
The response of LFGNM to cosmic ray modulations has been studied on a longer time scale for the period from September 2007 to September 2012 in conjunction with the response of the  conventional neutron monitors for the same period. The hourly data obtained from LFGNM is presented from September 2007 to September 2012 in figure 4. The 100\% count rate has been taken as an hourly average from 1st to 4th January 2008 as this corresponds to minimum of 23rd solar cycle \cite{Spaceweather}. The plot reveals an annual periodic/seasonal decrease averaging about 15\% commencing in the month of November and then reaching to its minimum in the month of February, and recovering back to its base line value in the month of April every year. This  annual variation in the LFGNM data has been consistently observed till September 2012. Importantly no significant flare activity or CME has been recorded on the sun during this period \cite{Spaceweather, NOAA} and therefore the observed  variation shown in figure 4 
cannot be attributed to any of the transient activity on the sun. The data during this period has been compared with the data obtained from  some  other conventional neutron monitors. It is observed that the data from  ESOI neutron monitor compares very well in terms of its profile matching with the data obtained from LFGNM (figure 4). The correlation coefficient obtained on the basis of hourly count rates for LFGNM and  ESOI neutron monitor turns out to be  0.67 during this period which gets further improved to 0.75 when calculated on the basis of 27 days moving averages (figure 5). It is worthwhile to mention here that  both these monitors are mid latitude and high altitude stations. LFGNM is operating at an altitude of 2743 m.a.s.l and the  geographic latitude of 34.07° N having geomagnetic cutoff rigidity of $\approx$ 11.4 GV, similarly  ESOI neutron monitor is operating at an altitude 2055 m.a.s.l and the geographic latitude of 33.3° N having geomagnetic cutoff rigidity of $\approx$ 10.8 GV. However, 
when the  data is 
compared with high latitude i.e. low rigidity 
neutron monitors like Inuvik, Irkutsk-1, Irkutsk-3, Kiel, Moscow and Norlisk the correlation coefficient becomes very poor and insignificant.

\section{Discussion and Conclusion}
LFGNM monitors a few MeV energy neutrons in the atmosphere with maximum efficiency of 3\% \cite{ShahPhd}. The profile of variation in the neutron count rate of LFGNM matches very well with that of recorded by low rigidity stations. The amplitude of decrease observed by LFGNM is comparable and consistent with that observed by the conventional neutron monitors, indicating that large scale changes do occur in the few MeV energy regime of secondary cosmic ray neutron component. This kind of response of LFGNM  makes it a proper candidate for extensive research in studying modulation effects comparatively in the lower energy regime of the secondary neutrons which hitherto has not been extensively studied. Therefore, it is emphasized that  LFGNM  responds as good  to solar modulation effects as any other conventional neutron monitor does. However, its response to such changes is large and indicates that solar modulation signatures can also very well and effectively be studied with the help of lead-free monitors. 
Furthermore, it also indicates that significant modulation effects do take place in the vicinity of the  energy regime where the LFGNM response is optimum. However, the periodic/seasonal modulation effect observed by LFGNM is unique in the sense that such effect has not been reported or seen in the data obtained from other conventional neutron monitors except  ESOI neutron monitor having 6NM64 configuration. It may be again emphasized here that these two neutron monitors have similarity in design to the extent that both of these monitors have comparable upper moderator thickness, otherwise the optimum sensitivity to the incident neutrons of these monitors is in different energy regimes. Therefore in the absence of any plausible explanation contributing to this periodic/seasonal modulation effect due to any extra-terrestial or heliospheric phenomena the role played by the variation in the station atmospheric temperature may not be excluded. At HARL, Gulmarg the absolute temperature variation of 36$^\circ$C in 
the local atmosphere from the summer season to the winter season seems to contribute this periodic/seasonal variation effect.  ESOI neutron monitor located at Mount Hermon, Israel witnesses same type of seasonal pattern and weather as that of HARL, Gulmarg, with well defined summer and winter seasons and exhibits similar kind of periodic/seasonal modulation effect in its recorded data. Therefore, the temperature variation in the local atmosphere seems to play a role in the periodic/seasonal modulation  effect being observed in the data of these two neutron monitor stations and this temperature-wave effect in the lower energy component of the secondary cosmic ray neutrons is nevertheless more pronounced. Whereas, through studies, it has been established \cite{Shah2010,Stoker2000} that environmental neutrons contribute  only less than 5\% to the overall countrate of the neutron monitors. Therefore, the authors feel that further study is required to conclusively establish and confirm the existence of this 
temperature-wave phenomenon.\\
\textbf{Acknowledgements}
\newline
 We are thankful to Emilio Segre, Athens and Tibet neutron monitor groups for making the data available on the internet. The authors are also thankful to R. Koul for continuously encouraging and motivating for pursuing this work, besides giving helpful suggestions for improving the contents of this paper.





\bibliographystyle{elsarticle-num}
\bibliography{<your-bib-database>}
 


\end{document}